\documentclass[
    10pt,
    superscriptaddress,
    aps,
    prx,
    twocolumn,
    nofootinbib,
    longbibliography
]{revtex4-2}

\usepackage{amsmath,amssymb,amsthm,amsfonts}
\usepackage{mathtools}
\usepackage{bm}
\usepackage{braket}
\usepackage{blkarray}
\usepackage{latexsym}

\usepackage{newtxtext,newtxmath}
\usepackage{anyfontsize}

\usepackage{graphicx}
\usepackage{epstopdf}
\usepackage{tikz}
\usepackage{subfigure}

\usepackage{multirow}
\usepackage{tabularx}
\usepackage{booktabs}
\usepackage{makecell}

\usepackage[version=4]{mhchem}
\usepackage{chemformula}
\usepackage{siunitx}

\usepackage[
    linesnumbered,
    lined,
    ruled,
    vlined
]{algorithm2e}

\usepackage{orcidlink}
\usepackage{appendix}
\usepackage{verbatim}
\usepackage[normalem]{ulem}
\usepackage{xurl}

\usepackage{lipsum}

\hypersetup{
    colorlinks=true,
    citecolor=blue,
    linkcolor=blue,
    urlcolor=blue,
    pdfborder={0 0 0}
}

\newcommand{\xrq}[1]{\textcolor{black}{#1}} 
\newcommand{\SVR}[1]{\textcolor{black}{#1}}

\begin{document}
\title{Resource-Efficient Counterdiabatic VQE in Symmetry-Reduced Active Spaces}

\author{Ruoqian Xu$^{\orcidlink{0009-0005-3242-9674}}$}
\affiliation{Quantum Advanced Research Center (QuARC), CSIC, 28049, Madrid, Spain}
\affiliation{Instituto de Ciencia de Materiales de Madrid (ICMM), CSIC, 28049, Madrid, Spain}
\affiliation{\mbox{Departamento de Física Teórica de la Materia Condensada, Universidad Autónoma de Madrid, 28049 Madrid, Spain}}

\author{Jialiang Tang$^{\orcidlink{0009-0001-6420-6492}}$}
\affiliation{Quantum Advanced Research Center (QuARC), CSIC, 28049, Madrid, Spain}
\affiliation{Instituto de Ciencia de Materiales de Madrid (ICMM), CSIC, 28049, Madrid, Spain}
\affiliation{\mbox{Departamento de Física Teórica de la Materia Condensada, Universidad Autónoma de Madrid, 28049 Madrid, Spain}}

\author{Sebastián V. Romero$^{\orcidlink{0000-0002-4675-4452}}$}
\affiliation{Quantum Advanced Research Center (QuARC), CSIC, 28049, Madrid, Spain}
\affiliation{Instituto de Ciencia de Materiales de Madrid (ICMM), CSIC, 28049, Madrid, Spain}
\affiliation{\mbox{Departamento de Física Teórica de la Materia Condensada, Universidad Autónoma de Madrid, 28049 Madrid, Spain}}

\author{Yue Ban$^{\orcidlink{0000-0003-1764-4470}}$}
\email{yue.ban@csic.es}
\affiliation{Quantum Advanced Research Center (QuARC), CSIC, 28049, Madrid, Spain}
\affiliation{Instituto de Ciencia de Materiales de Madrid (ICMM), CSIC, 28049, Madrid, Spain}

\author{Xi Chen$^{\orcidlink{0000-0003-4221-4288}}$}
\email{xi.chen@csic.es}
\affiliation{Quantum Advanced Research Center (QuARC), CSIC, 28049, Madrid, Spain}
\affiliation{Instituto de Ciencia de Materiales de Madrid (ICMM), CSIC, 28049, Madrid, Spain}

\date{\today}

\begin{abstract}
Accurate molecular ground-state calculations on quantum
hardware remain challenging due to substantial qubit
requirements and circuit complexity.
Here, we develop a resource-efficient framework combining
complete active-space (CAS) selection, molecular
symmetry-adapted qubit encoding, and counterdiabatic
(CD)-inspired variational ans\"atze, with Pauli-weight
truncation and adaptive operator selection applied
to operator pools derived from reduced CAS Hamiltonians.
Benchmarks on \ce{N2}, \ce{C2H4}, and \ce{C6H6}
using 7, 4, and 4 qubits, respectively, demonstrate
favorable accuracy-resource trade-offs relative to
hardware-efficient and symmetry-adapted UCCSD ans\"atze
within the selected active spaces.
Experiments on an IBM superconducting processor show
that Adaptive-CD-3 reduces transpiled two-qubit gate
counts by approximately 30\%-70\% relative to CD-3
across five benzene geometries while maintaining
comparable energy accuracy.
\end{abstract}

\maketitle

\section{Introduction}
\label{Sec1:intro}
 
Quantum computing offers a promising approach
to molecular electronic-structure calculations,
where accurately describing electron correlation
can become computationally demanding as system
size increases~\cite{RevModPhys.92.015003,10.1063/5.0173591}.
Classical methods, including configuration
interaction, coupled-cluster theory, and
density functional theory, have achieved
considerable success, but their accuracy
and computational cost depend strongly
on the electronic structure of the system~\cite{helgaker_2000,Shavitt_Bartlett_2009,
RevModPhys.79.291}.
Quantum algorithms such as quantum phase
estimation (QPE) and the variational quantum
eigensolver (VQE) provide alternative
approaches to molecular energy calculations~\cite{Peruzzo2014,McClean_2016,
PhysRevResearch.6.013221}.
Extensions of VQE have also been developed
for molecular excited states and low-energy
spectra~\cite{Higgott2019,Nakanishi2019,ollitraut2020quantum}. VQE can be implemented using relatively shallow parameterized quantum circuits, making it an attractive approach for near-term quantum processors. Nevertheless, its practical application remains limited by the qubit requirements of molecular Hamiltonians and the circuit resources needed to accurately represent correlated electronic states.

Reducing these requirements motivates
complementary strategies at the Hamiltonian
and circuit levels. Point-group symmetries
and conserved quantum numbers can be used
to restrict the Hamiltonian to the relevant
symmetry sector without altering its
spectrum within that sector~\cite{Setia_2020,PhysRevA.105.062452,
Picozzi_2023}.
Complete active-space (CAS) methods provide
an additional reduction by retaining selected
electrons and orbitals~\cite{roos2007complete,de2023complete}.
Unlike symmetry reduction, however,
active-space truncation introduces an
electronic-structure approximation.
At the circuit level, chemically motivated
ansätze such as unitary coupled-cluster
singles and doubles (UCCSD) can require
substantial resources, motivating more
compact constructions, including
$k$-UpCCGSD and low-rank unitary cluster
Jastrow approaches~\cite{lee2019generalized,WEIDMAN2024102105}.
Adaptive algorithms, notably ADAPT-VQE,
further reduce unnecessary circuit
operators through gradient-based selection~\cite{Grimsley2019,Tang2021,Yordanov2021}.

Counterdiabatic (CD) driving, originally
developed to suppress nonadiabatic transitions~\cite{DemirplakRice2003,Berry2009},
provides another physics-informed route
to quantum-state preparation.
Within the framework of shortcuts to
adiabaticity (STA), approximate adiabatic
gauge potentials (AGPs) can be constructed
using variational methods and nested
Hamiltonian commutators~\cite{Sels2017,Claeys2019}.
These approaches have motivated
CD-assisted quantum algorithms and
variational circuit constructions~\cite{Hegade2021, PhysRevApplied.21.034033,Tang2024}.
In particular, AGP-derived Pauli operators
have been employed as independent
variational generators in the adiabatic
gauge ansatz and its reduced implementation~\cite{FerreiroVelez2026}.
Counterdiabatic ADAPT-VQE has also combined
AGP-derived operator pools with adaptive
gradient-based selection~\cite{Tancara2026}.
Despite these advances, extending CD-inspired
variational approaches to more complex molecular
systems remains challenging because the
increasing size of molecular Hamiltonians
can lead to large AGP-derived operator pools
and costly circuit implementations. 
This motivates constructing CD-inspired
variational operators directly from
symmetry-reduced active-space Hamiltonians
and examining the resulting trade-off
between energy accuracy and circuit resources.
 
\begin{figure}[t]
    \centering
\includegraphics[width=\linewidth]{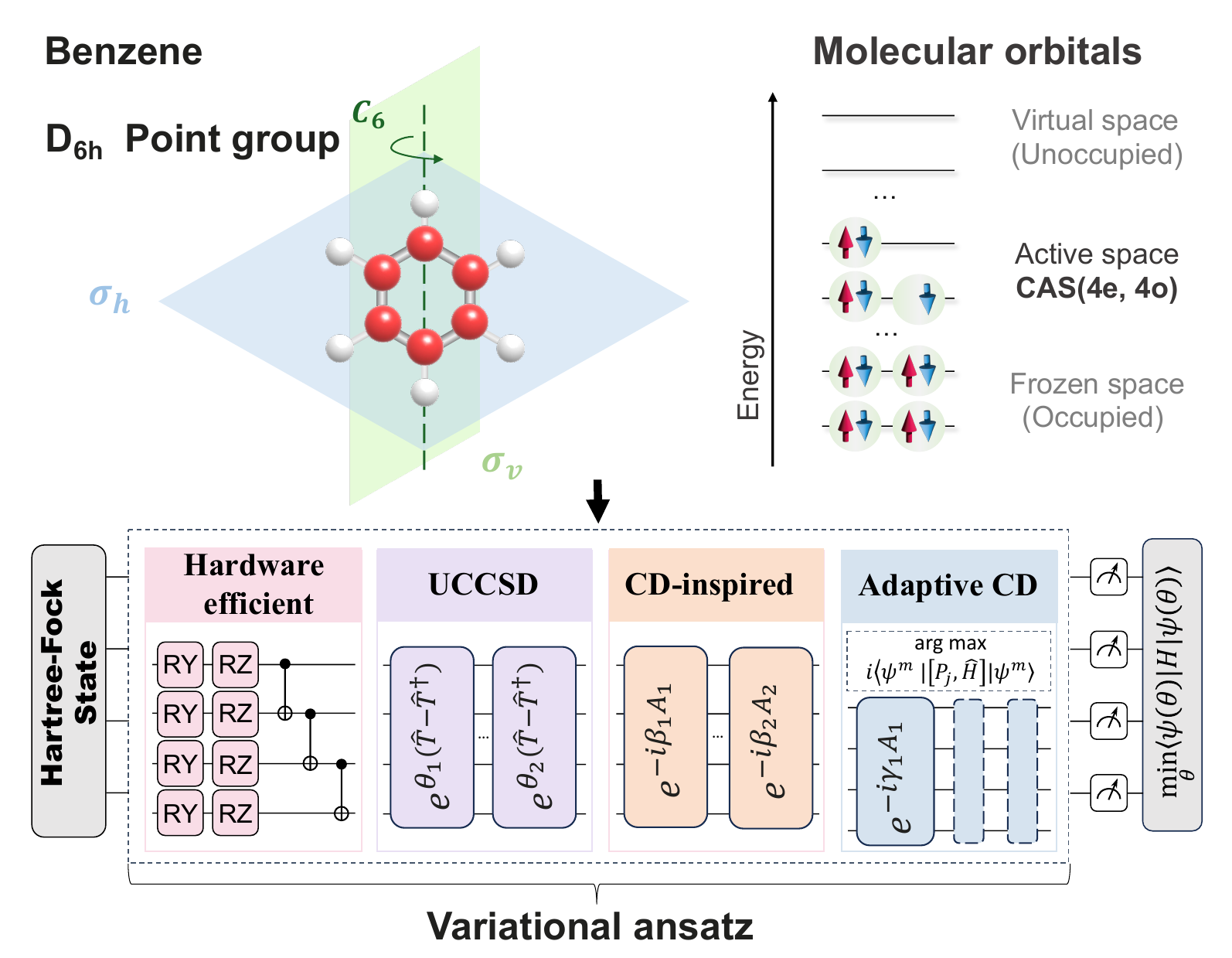}\vspace{-2mm}%
    \caption{Schematic illustration of the framework
combining CAS selection, molecular
symmetry-adapted encoding, and
variational circuit construction.
The lower panel illustrates four
ansatz families: HEA, SAE-UCCSD,
fixed CD-inspired, and adaptive
CD-inspired circuits.
All circuits are initialized from
encoded HF reference states.}
    \label{fig: cir}
\end{figure}

In this work, we investigate the integration
of CAS Hamiltonians, symmetry-adapted
qubit encoding, and CD-inspired variational
circuits, as illustrated in
Fig.~\ref{fig: cir}.
We employ fixed-orbital complete active-space configuration interaction (CASCI) calculations
to construct the molecular Hamiltonians
and use point-group symmetry reduction to obtain compact qubit representations. Approximate AGPs are then used to generate variational operator pools directly from the reduced spaces. We examine both fixed Pauli-weight truncation and adaptive operator selection for \ce{C6H6}, \ce{C2H4}, and \ce{N2}. By comparing these constructions with
hardware-efficient ansatz (HEA) and UCCSD implemented with symmetry-adapted encoding (SAE-UCCSD), we assess variational accuracy, optimization convergence, and circuit resources over a range of molecular geometries. We further investigate the impact of noise and implement our circuits on IBM hardware. Our results underscore the potential of combining molecular Hamiltonian reduction and CD-inspired operator selection to improve accuracy-resource trade-offs for chemically complex molecular systems.

The remainder of this paper is organized
as follows. Sec.~\ref{sec:molecular_model}
introduces the molecular Hamiltonians,
active-space approximation, and
symmetry-adapted qubit encoding.
Sec.~\ref{sec:VQA} describes the
variational ansätze.
Numerical benchmarks and quantum hardware
results are presented in
Secs.~\ref{sec:simu_results}
and~\ref{sec: exp}, respectively.
Sec.~\ref{sec: conclu} concludes
with a discussion of the results
and future perspectives.

\section{Hamiltonians and Qubit Encoding}
\label{sec:molecular_model}

\subsection{Electronic-structure Hamiltonian}
\label{sec:electronic_hamiltonian}

Within the Born-Oppenheimer approximation, the
nuclear coordinates $\mathbf{R}$ are treated as
fixed parameters~\cite{helgaker_2000,szabo1996modern}.
In atomic units, the electronic Hamiltonian is
given by
\begin{equation}
\label{eq:h_e}
\hat{H}_{\mathrm{el}}(\mathbf{R})
=
-\frac{1}{2}\sum_{i=1}^{N}\nabla_i^2
-\sum_{i=1}^{N}\sum_{A=1}^{N'}
\frac{Z_A}{r_{iA}}
+\sum_{i<j}^{N}\frac{1}{r_{ij}},
\end{equation}
where $N$ and $N'$ denote the
numbers of electrons and nuclei, respectively.
The total molecular energy also includes the
nuclear repulsion contribution,
$E_{\mathrm{NN}}(\mathbf{R})
=\sum_{A<B}Z_AZ_B/R_{AB}$.

Expanding the electronic Hamiltonian on a finite
orthonormal spin-orbital basis and including
the nuclear repulsion energy yield the
second-quantized molecular Hamiltonian
~\cite{helgaker_2000,RevModPhys.92.015003},
\begin{equation}
\label{eq:second_quantized}
\hat{H}
=
E_{\mathrm{NN}}
+\sum_{pq}h_{pq}\hat{a}_p^\dagger\hat{a}_q
+\frac{1}{2}\sum_{pqrs}g_{pqrs}
\hat{a}_p^\dagger\hat{a}_q^\dagger
\hat{a}_s\hat{a}_r,
\end{equation}
where $\hat{a}_p^\dagger$ and $\hat{a}_p$
are fermionic creation and annihilation
operators. The one- and two-electron
integrals are defined as
\begin{align}\label{eq:hpq}
h_{pq}
&=
\int \text{d}x\,\chi_p^*(x)
\left[
-\frac{\nabla^2}{2}
-\sum_A\frac{Z_A}{r_A}
\right]\chi_q(x),
\\ \label{eq:gpqrs}
g_{pqrs}
&=
\iint \text{d}x_1\text{d}x_2\,
\frac{
\chi_p^*(x_1)\chi_q^*(x_2)
\chi_s(x_2)\chi_r(x_1)
}{r_{12}},
\end{align}
where $\chi_p(x)$ denotes the $p$-th spin orbital,
the integral over $x$ includes spatial and spin coordinates, i.e., $\int \text{d}x \equiv \sum_{\sigma=\alpha,\beta} \int_{\mathbb R^3}\text{d}^3\mathbf r$ and $g_{pqrs}$ denotes the
non-antisymmetrized two-electron integral.

The Hartree-Fock (HF) approximation provides
a single-determinant reference state and a set
of molecular orbitals~\cite{szabo1996modern}.
Electronic correlations beyond the mean-field
description are subsequently treated within
selected active spaces.

\subsection{CAS approximation}
\label{sec:active_space}

The CAS approach provides
a systematic framework for treating electron
correlation within a selected set of molecular
orbitals~\cite{Roos1980,helgaker_2000,roos2007complete}.
In a $\mathrm{CAS}(N_{\mathrm{e}},M_{\mathrm{o}})$
description, $N_{\mathrm{e}}$ active electrons
are distributed among $M_{\mathrm{o}}$
active spatial orbitals, while the remaining
orbitals are classified as doubly occupied
inactive orbitals or unoccupied external
orbitals. Within the selected active space,
all electronic configurations consistent
with the prescribed electron number and
symmetry constraints are retained.

In this work, we employ a CASCI approach,
in which the molecular orbitals obtained from
HF calculations remain fixed during the
configuration interaction calculation.
The active-space Hamiltonian includes the
interactions among active electrons, the
effective one-body contributions from inactive
orbitals, and the corresponding constant
core energy. The CASCI reference energies
are obtained by diagonalizing this Hamiltonian
within the selected active space.
Active-space
truncation introduces an approximation
whose accuracy depends on the choice of
active orbitals and the electronic
correlations retained.

\begin{figure}[t]
\centering
\includegraphics[width=\linewidth]{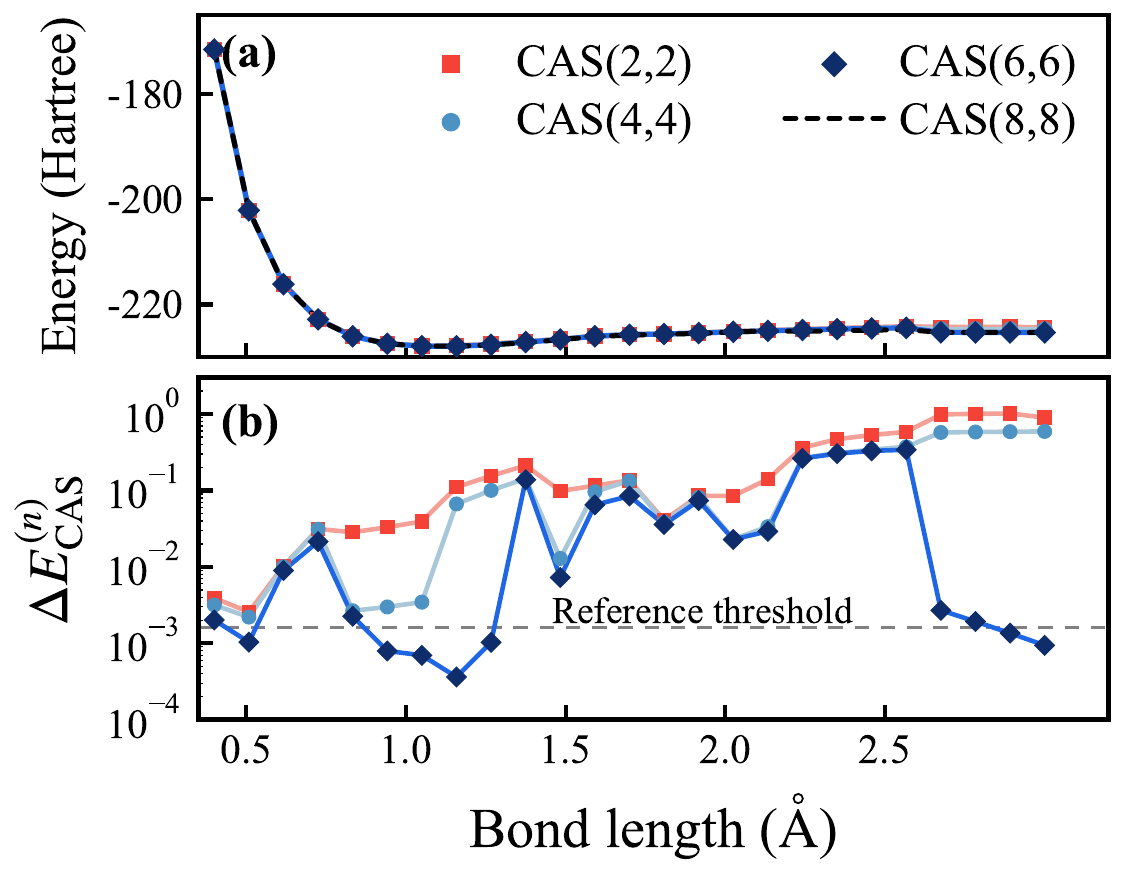}\vspace{-3mm}
\caption{
{Potential-energy curves and active-space energy
deviations for
\ce{C6H6} near equilibrium bond distance.}
(a) Potential energy curves obtained using
$\mathrm{CAS}(2,2)$, $\mathrm{CAS}(4,4)$,
and $\mathrm{CAS}(6,6)$, compared with the
$\mathrm{CAS}(8,8)$ reference.
(b) Corresponding energy deviations $\Delta E_{\mathrm{CAS}}^{(n)}$ [Eq.~\eqref{eq:cas_error}] as functions of bond length. The horizontal dashed line marks an energy-deviation reference threshold of $1.6$~mHa.}
\label{fig:cas}
\end{figure}

To investigate the accuracy--resource trade-off,
we consider a hierarchy of active spaces,
$\mathrm{CAS}(2,2)$, $\mathrm{CAS}(4,4)$,
$\mathrm{CAS}(6,6)$, and $\mathrm{CAS}(8,8)$.
Fig.~\ref{fig:cas} compares the corresponding
potential-energy curves of \ce{C6H6}, using the
$\mathrm{CAS}(8,8)$ result as a reference. The calculated energies generally approach
this reference as the active space increases.
The deviations are particularly pronounced
away from equilibrium, reflecting the
additional electronic correlations represented
by the larger active spaces.
Similar active-space comparisons for
\ce{C2H4} and \ce{N2} are presented in \xrq{Section S1 of the Supporting Information.}

\subsection{Symmetry-adapted encoding}
\label{sec:group_symmetry}

Molecular point-group symmetries provide an additional
means of reducing the computational
representation without introducing further
approximations within a selected symmetry
sector. For a molecular Hamiltonian invariant
under a point group $\mathcal{G}$, the electronic
Hilbert space decomposes into invariant
subspaces associated with the irreducible
representations of $\mathcal{G}$.
In a symmetry-adapted basis, the Hamiltonian
takes a block-diagonal form, schematically
written as $\hat{H} = \bigoplus_{\Gamma}\hat{H}_{\Gamma}$, 
where $\Gamma$ labels the symmetry sectors.

At their equilibrium geometries, the molecules
considered in this work belong to the point
groups $D_{\infty h}$ for \ce{N2},
$D_{2h}$ for \ce{C2H4}, and
$D_{6h}$ for \ce{C6H6}.
These correspond to a linear molecule with an
infinite-fold rotation axis and perpendicular
twofold axes, a planar molecule with three
mutually perpendicular twofold axes, and a
hexagonal planar molecule with a sixfold axis
perpendicular to the molecular plane and six
in-plane twofold axes, respectively. All three
possess inversion symmetry. For \ce{N2}, the
finite Abelian subgroup $D_{2h}$ is commonly
employed in electronic-structure calculations
~\cite{Cotton1990,10.1063/5.0006074}.

To exploit conserved symmetries, we consider
independent, mutually commuting
$\mathbb{Z}_2$ symmetry operators
$\hat{S}_{\mu}$, which may arise from
molecular spatial symmetries or conserved
fermionic parities. These operators satisfy $\hat{S}_{\mu}^{2}=\hat{I}$, $[\hat{H},\hat{S}_{\mu}]=0$, and $[\hat{S}_{\mu},\hat{S}_{\nu}]=0$.
Each symmetry sector is specified by a set
of eigenvalues $s_{\mu}=\pm1$, such that $\hat{S}_{\mu}|\Psi\rangle = s_{\mu}|\Psi\rangle$.
Restricting the Hamiltonian to the sector
containing the target electronic state
preserves its spectrum within that sector
while reducing the computational dimension.

\subsection{Fermion-to-qubit mapping}
\label{sec:qubit_mapping}

To implement the active-space Hamiltonian
on a quantum processor, fermionic creation
and annihilation operators are mapped onto
qubit operators. Here, we employ the
Jordan--Wigner (JW) transformation~\cite{Jordan1928,Seeley2012},

\color{black}
\begin{align}
    \hat{a}_j^\dagger
\mapsto
\frac{1}{2}(X_j-iY_j)
\prod_{k<j}Z_k,
\quad
\hat{a}_j
\mapsto
\frac{1}{2}(X_j+iY_j)
\prod_{k<j}Z_k,
\end{align}\color{black}
where $X_j$, $Y_j$, and $Z_j$ denote
Pauli operators acting on qubit $j$. {For an active space containing $M_{\mathrm{o}}$ spatial
orbitals, the JW transformation yields a Hamiltonian
acting on $2M_{\mathrm{o}}$ qubits, expressed as a weighted sum of Pauli strings.}

The symmetry constraints discussed in
Sec.~\ref{sec:group_symmetry} can then
be applied to the mapped Hamiltonian
to restrict the calculation to the
symmetry sector containing the target
electronic state
~\cite{Setia_2020,PhysRevA.105.062452,Picozzi_2023}. By eliminating $r_{\mathrm{sym}}$ independent
binary degrees of freedom through symmetry encoding,
the number of retained qubits becomes $n_q = 2M_{\mathrm{o}}-r_{\mathrm{sym}}$,
where $r_{\mathrm{sym}}$ is the number of independent
symmetry constraints used in the encoding. The resulting symmetry-reduced
Hamiltonian is expressed as $\hat{H}_{q} = \sum_{\ell=1}^{K}c_{\ell}P_{\ell}$,
where
$P_{\ell}\in\{I,X,Y,Z\}^{\otimes n_q}$
acts on the $n_q$ retained qubits,
$K$ denotes the number of nonzero
Pauli terms, and $c_{\ell}\in\mathbb{R}$. The symmetry-reduced Hamiltonian
$\hat{H}_{q}$ provides the starting
point for the CD-inspired
variational ansatz constructions
developed in the following section.

\begin{table}[t]
\centering
\caption{
Qubit counts, numbers of Pauli terms, and active-space
energy deviations for symmetry-reduced \ce{C6H6}
Hamiltonians. Energy deviations are evaluated relative
to $\mathrm{CAS}(8,8)$ using Eq.~\eqref{eq:cas_error}
and reported in Hartree. {
The variation in qubit counts across molecular
geometries is associated with changes in the
irreducible representations of the selected
active orbitals, which enter the construction
of the symmetry-adapted encoding.
}
}
\begin{ruledtabular}
\begin{tabular}{lcccc}
Active space & Qubits & Pauli terms
& Median $\Delta E$ & Max. $\Delta E$ \\ \midrule
$\mathrm{CAS}(2,2)$ & 1
& 3 & $0.131$ & $1.01$ \\
$\mathrm{CAS}(4,4)$ & 3--4
& 20--46 & $0.100$ & $0.593$ \\
$\mathrm{CAS}(6,6)$ & 7--8
& 224--466 & $0.00897$ & $0.266$ \\
$\mathrm{CAS}(8,8)$ & 11--12
& 665--1473 & --- & --- \\
\end{tabular}
\end{ruledtabular}
\label{tab:c6h6_cas_resources}
\end{table}
To quantify the effect of active-space
truncation, we define the energy deviation
for $\mathrm{CAS}(n,n)$, with $n=2,4,6,8$,
relative to the $\mathrm{CAS}(8,8)$ reference:
\begin{equation}
\label{eq:cas_error}
\Delta E_{\mathrm{CAS}}^{(n)}(R)
=
\left|
E_{\mathrm{CASCI}(n,n)}(R)
-
E_{\mathrm{CASCI}(8,8)}(R)
\right|,
\end{equation}
where $R$ denotes the varied bond distance,
and all energies are evaluated at the same
molecular geometry and using the same
one-electron basis. Table~\ref{tab:c6h6_cas_resources}
summarizes the corresponding energy
deviations and quantum resources after
symmetry reduction.
Increasing the active space from
$\mathrm{CAS}(4,4)$ to $\mathrm{CAS}(6,6)$
reduces the median energy deviation
from approximately $100$ to $9$~mHa,
while increasing the qubit requirements
and the number of Pauli terms.
The $\mathrm{CAS}(8,8)$ Hamiltonians
require $11$--$12$ qubits and contain
up to $1473$ Pauli terms.
The $\mathrm{CAS}(8,8)$ energies serve
as references within the investigated
active-space hierarchy rather than
full-space FCI benchmarks.
The corresponding deviations therefore
quantify active-space convergence,
not the accuracy relative to the
full molecular Hamiltonian.

\begin{figure}[t]
\centering
\includegraphics[width=\linewidth]{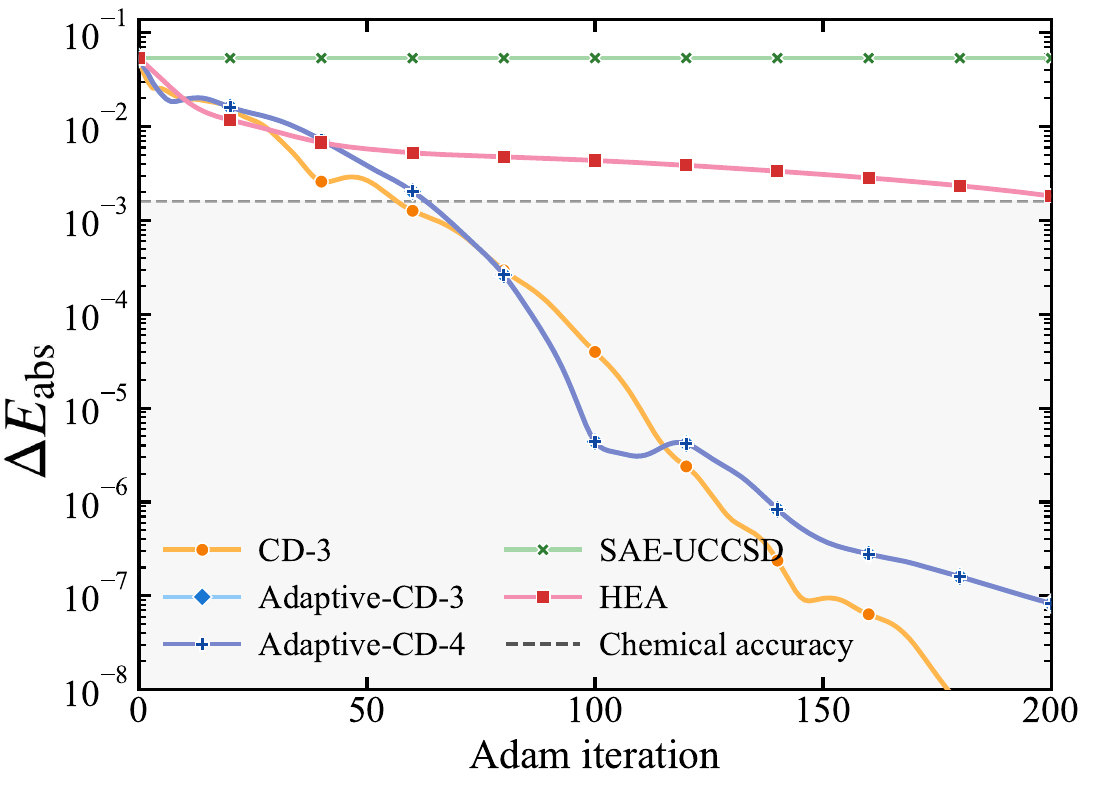}\vspace{-3mm}
\caption{
Convergence of the absolute energy
deviation $\Delta E_{\mathrm{abs}}$~[Eq.~\eqref{eq:ideal_delta_E}] as a
function of Adam optimization iterations
for \ce{C6H6} [$\mathrm{CAS}(4,4)$]
at the equilibrium bond length $R$(C--H) = 1.09 $\mathrm{\mathring{A}}$. 
The compared ansätze
are CD-3, Adaptive-CD-3, Adaptive-CD-4,
SAE-UCCSD, and HEA. Adam optimization
uses a learning rate of $0.01$,
with initial variational parameters
set to zero. The dashed line marks
the $1.6$~mHa accuracy benchmark
relative to the active-space
FCI reference.
}
\label{fig: adam iteration}
\end{figure}

\section{Variational quantum algorithms}
\label{sec:VQA}

Variational quantum algorithms (VQAs) combine
parameterized quantum circuits with classical
optimization to approximate molecular ground states
on quantum processors
~\cite{McClean_2016,PhysRevLett.120.110501,
PhysRevX.8.011044}.
Starting from a reference state
$|\psi_{\mathrm{ref}}\rangle$, a parameterized
circuit $U(\boldsymbol{\theta})$ prepares a
trial state whose energy is minimized according to
\begin{equation}
\label{eq:vqe_energy}
E(\boldsymbol{\theta})
=
\langle\psi_{\mathrm{ref}}|
U^\dagger(\boldsymbol{\theta})
\hat{H}_q
U(\boldsymbol{\theta})
|\psi_{\mathrm{ref}}\rangle,
\end{equation}
where $\hat{H}_q$ is the symmetry-reduced
molecular Hamiltonian defined in
Sec.~\ref{sec:qubit_mapping}.
The choice of ansatz determines the
accessible variational states and
the circuit resources required
for their preparation.

Our CD-inspired approach constructs variational
generators from approximate AGPs associated with the molecular
Hamiltonian. We investigate two complementary
strategies: a fixed ansatz constructed from
Pauli operators with bounded weight and an
adaptive ansatz that selects operators
according to their energy gradients. To assess their performance, we consider
two reference ansätze: a hardware-efficient
ansatz (HEA) and UCCSD with symmetry-adapted
encoding (SAE-UCCSD).  The HEA consists of parameterized single-qubit
rotations and entangling gates, whereas the
SAE-UCCSD ansatz employs excitation operators
motivated by electronic-structure theory~\cite{romero2018strategies}. Their circuit constructions are described
in \xrq{Section S2 of the Supporting Information.}

\subsection{CD-inspired variational circuits}
\label{sec:CD_ansatz}

We consider an adiabatic Hamiltonian with the interpolation
$\hat H_a(\lambda)
=
(1-\lambda)\hat H_i+\lambda\hat H_q$,
where $\lambda(t)\in[0,1]$ $\forall t\in[0,T]$ is a switching function satisfying $\lambda(0)=0$ and $\lambda(T)=1$, $\hat H_q$
is the target molecular Hamiltonian,
and $\hat H_i$ is a reference Hamiltonian
whose ground state is the encoded
HF state. This state provides a suitable initial reference while requiring minimal resources for its preparation, involving only local gates applied to the zero state.
For an HF bitstring
$|\Phi_{\mathrm{HF}}\rangle
=|b_0b_1\cdots b_{n_q-1}\rangle$,
we choose
$\hat H_i
=
\sum_{j=0}^{n_q-1}(2b_j-1)Z_j$,
where $b_j\in\{0,1\}$.
This choice makes $|\Phi_{\mathrm{HF}}\rangle$
the unique ground state of $\hat H_i$.
For the four-qubit benzene model, the encoded
HF state is $|\Phi_{\mathrm{HF}}\rangle=|1000\rangle$,
yielding the reference Hamiltonian
$\hat H_i=Z_0-Z_1-Z_2-Z_3$.

Developed within the framework of shortcuts to adiabaticity (STA), CD driving~\cite{DemirplakRice2003,Berry2009}
suppresses nonadiabatic transitions
through an auxiliary Hamiltonian
determined by the AGP, which is included in the adiabatic Hamiltonian as $H_\mathrm{CD}(\lambda) = H_a(\lambda) + \dot{\lambda}\hat{\mathcal{A}}_\lambda$, with $\dot{\lambda}(0)=\dot{\lambda}(T)=0$ and $\hat{\mathcal{A}}_\lambda$ the exact AGP. Since its exact form generally involves 
nonlocal operators, we adopt the nested-commutator approximation~\cite{Claeys2019},%
\begin{equation}
\label{eq:nc_agp}
\hat{\mathcal A}_{\lambda}^{(\ell)}
=
i\sum_{k=1}^{\ell}\alpha_k(\lambda)
\underbrace{
[\hat H_a,[\hat H_a,\ldots,
[\hat H_a}_{2k-1},
\partial_\lambda\hat H_a]
\ldots]],
\end{equation}
where $\ell$ denotes the expansion order and $\alpha_k(\lambda)$ are the corresponding coefficients.
We construct the CD operator pool
$\mathcal P_{\mathrm{CD}}$ by collecting
distinct Pauli strings generated by the
first-order nested commutator ($\ell=1$),
retaining terms whose coefficient
magnitudes exceed $10^{-4}$.
Then, these operators are employed as variational
generators with independently optimized
rotation angles, rather than being used
to implement CD evolution
directly \cite{FerreiroVelez2026}.

Starting from the HF reference state $|\Phi_{\mathrm{HF}}\rangle$, the resulting CD-inspired variational ansatz is expressed as
\begin{equation}
\label{eq:cd_ansatz}
|\psi_{\mathrm{CD}}(\boldsymbol{\theta})\rangle
=
\prod_{l=1}^{D}
\prod_{j=1}^{J}
e^{-i\theta_{l,j}P_j}
|\Phi_{\mathrm{HF}}\rangle,
\end{equation}
where $P_j\in\mathcal P_{\mathrm{CD}}$, $J=|\mathcal{P}_\text{CD}|$ is the number of operators in the pool, and $D$ is the number of circuit layers. Each Pauli rotation has an independent variational parameter. Throughout this work, we set $D=1$ and use a fixed operator ordering.

\begin{figure*}[t]
\centering
\includegraphics[width=\linewidth]{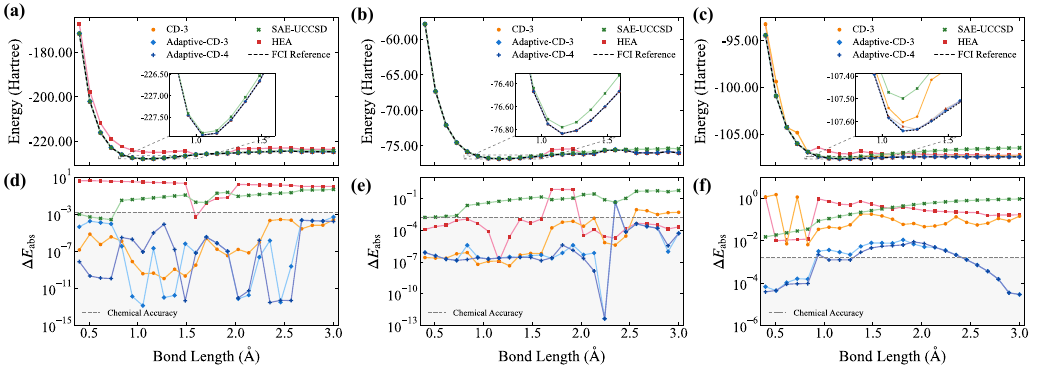}\vspace{-4mm}
\caption{
Potential-energy curves and corresponding
absolute energy deviations for the
three molecular systems.
(a)--(c) Potential-energy curves of
\ce{C6H6} [$\mathrm{CAS}(4,4)$],
\ce{C2H4} [$\mathrm{CAS}(4,4)$],
and \ce{N2} [$\mathrm{CAS}(6,6)$],
represented using 4, 4, and 7 qubits,
respectively, after symmetry reduction.
The indicated C--H, C--H, and N--N
bond distances are varied while the
remaining geometric parameters are
held fixed. Black dashed curves show
the corresponding active-space FCI
reference energies obtained using
PySCF~\cite{10.1063/5.0006074}.
(d)--(f) Absolute energy deviations~[Eq.~\eqref{eq:ideal_delta_E}] for the five
ansätze. The gray shaded region
indicates deviations below
$1.6$~mHa relative to the
active-space FCI reference,
and black-edged markers identify
the equilibrium geometries.
}
\label{fig: 3moles}
\end{figure*}

To reduce circuit complexity, we further truncate the
CD operator pool according to the Pauli weight $w(P_j)$,
defined as the number of nonidentity single-qubit
Pauli factors in the Pauli string $P_j$.
Introducing $w_{\max}$ as the maximum allowed Pauli
weight, we define the truncated operator pool as
\begin{equation}
\label{eq:cd_truncated_pool}
\mathcal{P}_{\mathrm{CD}}^{(w_{\max})}
=
\left\{
P_j\in\mathcal{P}_{\mathrm{CD}}
\,\middle|\,
w(P_j)\leq w_{\max}
\right\}.
\end{equation}
In particular, the CD-3 ansatz corresponds to
$w_{\max}=3$, retaining only Pauli strings acting nontrivially on at most three qubits.
This truncation reduces the implementation cost of the corresponding Pauli rotations.
Additional truncation levels are examined \xrq{in Section S3 of the
Supporting Information}.

\subsection{Adaptive CD ansatz}
\label{sec:adaptive_cd}

To further reduce circuit complexity, we combine
the truncated CD operator pool with the adaptive
selection strategy of ADAPT-VQE~\cite{Grimsley2019}, as previously applied in molecular simulation~\cite{Tancara2026}. Starting from the HF reference state,
the ansatz is iteratively expanded by appending
Pauli rotations selected from the truncated pool.
After $m$ iterations, the variational state reads
\begin{equation}
\label{eq:adaptive_cd_ansatz}
|\psi^{(m)}\rangle
=
e^{-i\theta_m P_{j_m}}
\cdots
e^{-i\theta_1 P_{j_1}}
|\Phi_{\mathrm{HF}}\rangle,
\end{equation}
where
$P_{j_k}\in\mathcal{P}_{\mathrm{CD}}^{(w_{\max})}$.
At each iteration, the candidate generators are
ranked by the magnitude of the energy gradient,
\begin{equation}
\label{eq:adaptive_gradient}
g_j^{(m)}
=
\left|
\langle\psi^{(m)}|
[P_j,\hat H_q]
|\psi^{(m)}\rangle
\right|.
\end{equation}
At each adaptive step, the operator with
the largest absolute energy gradient is
appended to the circuit, followed by
reoptimization of all variational
parameters.
The adaptive construction terminates when
the Euclidean norm of the candidate-gradient
vector falls below $10^{-3}$ or when
50 operators have been selected.

We consider Adaptive-CD-3 and Adaptive-CD-4,
corresponding to maximum Pauli weights of
$w_{\max}=3$ and $4$, respectively.
Unlike the fixed CD ansatz, these circuits
contain only the adaptively selected generators,
allowing the ansatz to grow according to the
energy response of the variational state.

\section{Performance and Noise Robustness}
\label{sec:simu_results}

We benchmark the CD-inspired ansatz
against HEA and
SAE-UCCSD for \ce{N2},
\ce{C2H4}, and \ce{C6H6} using the MindQuantum
simulation framework~\cite{mindquantum}. The active-space Hamiltonians are constructed
from HF molecular orbitals obtained
using PySCF~\cite{10.1063/5.0006074}
with the STO-3G basis set to compute the integrals in Eqs.~\eqref{eq:hpq}-\eqref{eq:gpqrs}. Active orbitals are selected according to the default HF orbital ordering, without further orbital optimization. We employ $\mathrm{CAS}(6,6)$ for \ce{N2} and $\mathrm{CAS}(4,4)$ for both \ce{C2H4} and \ce{C6H6}, yielding 7, 4, and 4 qubits, respectively, after symmetry reduction.
For \ce{C6H6}, the resulting active space
contains four frontier molecular orbitals
rather than the complete six-orbital $\pi$ manifold. 

\begin{figure*}[t]
\centering
\includegraphics[width=\linewidth]{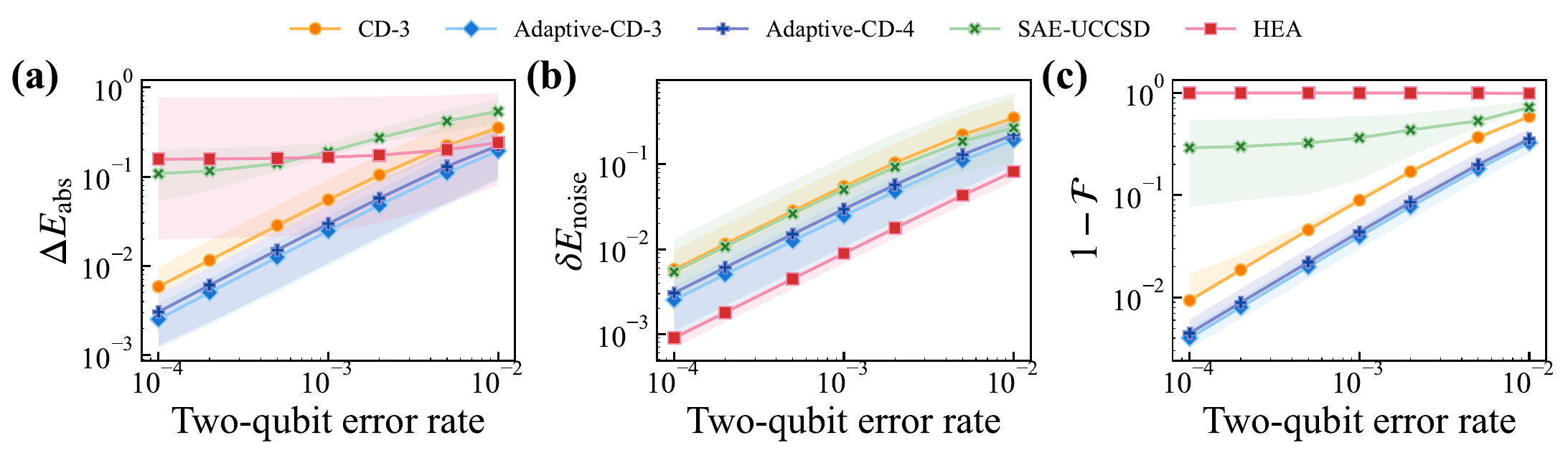}\vspace{-2mm}%
\caption{
(a) Absolute energy deviation~$\Delta E_{\mathrm{abs}}~$[Eq.~\eqref{eq:ideal_delta_E}] as a function of
the two-qubit error rate $p_{2q}$.
(b) Noise-induced energy deviation~$\delta E_{\mathrm{noise}}~$[Eq.
~\eqref{eq:noise_error}] as a function of
$p_{2q}$.
(c) Ground-state infidelity~$1-\mathcal{F}~$[Eq.~\eqref{eq:infidelity}] as a
function of $p_{2q}$.
Solid lines in (a)--(c) represent the medians
over the investigated molecular geometries,
while the shaded regions indicate the
interquartile ranges.}
\label{fig:robustness}
\end{figure*}

\subsection{Ideal-circuit benchmarks}

Variational energies are benchmarked against
the exact ground-state energies of the
corresponding active-space Hamiltonians.
We define the absolute energy deviation as
\begin{equation}
\label{eq:ideal_delta_E}
\Delta E_{\mathrm{abs}}(R)
=
|E(R)-E_{\mathrm{FCI}}(R)|,
\end{equation}
where $E(R)$ denotes the ideal variational
energy or its noisy or hardware estimate,
as appropriate, and $E_{\mathrm{FCI}}(R)$
is the exact ground-state energy of the
corresponding symmetry-reduced
active-space Hamiltonian.

We first examine the parameter-optimization
convergence of five variational ansätze for
\ce{C6H6} [$\mathrm{CAS}(4,4)$]
at its equilibrium geometry.
The compared methods are CD-3,
Adaptive-CD-3, Adaptive-CD-4, HEA,
and SAE-UCCSD.
For the adaptive ansätze, the operator
sequences are determined beforehand.
All circuit structures are then kept fixed,
and their variational parameters are
initialized to zero.
Each ansatz is optimized using Adam
with a learning rate of $0.01$
for up to 200 iterations.
The adaptive operator-selection and
intermediate optimization costs are
not included in the iteration count.
As shown in Fig.~\ref{fig: adam iteration},
the CD-inspired ansätze exhibit faster
energy convergence than HEA and SAE-UCCSD
during this fixed-circuit optimization.
CD-3 and the adaptive variants reach the
$1.6$~mHa accuracy threshold within the
first 100 iterations, whereas HEA approaches
this threshold more gradually.
SAE-UCCSD retains a larger energy deviation
after 200 iterations.
These results suggest that AGP-derived
variational generators facilitate
parameter optimization for the present
symmetry-reduced molecular Hamiltonian.

We next compare the potential-energy
curves over the investigated molecular
geometries after 200 optimization
iterations. Figs.~\ref{fig: 3moles}(a)--(c)
show the variational energies of
\ce{C6H6}, \ce{C2H4}, and \ce{N2},
respectively, while
Figs.~\ref{fig: 3moles}(d)--(f)
present the corresponding absolute
deviations $\Delta E_{\mathrm{abs}}$ from the active-space
FCI references. All five ansätze
reproduce the overall energy profiles,
although their quantitative accuracy
varies with molecular geometry. For \ce{C6H6}, both fixed and adaptive
CD-inspired ansätze maintain relatively
small energy deviations over the
investigated bond-coordinate range,
with the adaptive variants reaching
the $1.6$~mHa benchmark at many
geometries. The performance of CD-3
suggests that the retained low-weight
AGP-derived operators already provide
an effective variational representation
within the four-qubit
$\mathrm{CAS}(4,4)$ model.

For \ce{C2H4}, the adaptive CD ansatz
generally yields smaller energy deviations
than fixed CD-3, particularly at larger
bond distances where the accuracy of
the fixed ansatz deteriorates.
The HEA and SAE-UCCSD benchmarks
also exhibit pronounced
geometry-dependent errors.
These results suggest that adaptive
selection from the CD operator pool
can improve the variational
representation as the molecular
geometry changes.
The \ce{N2} dissociation curve provides
a more challenging benchmark.
Although all methods reproduce the
overall energy profile, {several ansätze exhibit substantial energy deviations in the stretched-bond regime, while the adaptive CD constructions maintain higher accuracy at selected bond lengths.}

Adaptive-CD-3 and Adaptive-CD-4
improve the accuracy over selected
geometries, but the $1.6$~mHa
benchmark is not maintained
throughout the entire curve.
This behavior is consistent with
the increasing importance of
static correlation during bond
dissociation, where a restricted
operator pool may become insufficient
to describe the ground state
accurately. Thus, the geometry-dependent performance highlights
the trade-off between operator-pool complexity
and variational accuracy, particularly in
bond-stretched configurations.

\subsection{Noise sensitivity and circuit resources}
\label{sec:noisy_simulation}

We next examine the effect of gate noise on the
optimized variational circuits. For each ansatz,
the circuit structure and variational parameters
are first determined in noiseless simulations
and subsequently kept fixed during the noisy
calculations. This protocol isolates the effect
of gate noise from that of noisy variational
optimization. We employ gate-dependent depolarizing channels,
with single- and two-qubit noise parameters
related by $p_{1q}=0.1p_{2q}$. 
We characterize the performance under
gate noise using three complementary
measures: the absolute energy deviation,
the noise-induced energy deviation,
and the ground-state infidelity.
For the absolute energy deviation,
we set $E(R)=E_{\mathrm{noisy}}(R)$
in Eq.~\eqref{eq:ideal_delta_E}.
The remaining two measures are defined as
\begin{align}
\delta E_{\mathrm{noise}}(R)
&=
|E_{\mathrm{noisy}}(R)-E_{\mathrm{ideal}}(R)|,
\label{eq:noise_error}\\
1-\mathcal{F}
&=
1-\langle\psi_0|
\rho_{\mathrm{noisy}}|\psi_0\rangle,
\label{eq:infidelity}
\end{align}
where $E_{\mathrm{ideal}}$ is the noiseless
variational energy, $\rho_{\mathrm{noisy}}$
is the noisy output state, and
$|\psi_0\rangle$ is the exact ground
state of the corresponding active-space
Hamiltonian.
The noise-induced energy deviation
quantifies the magnitude of the energy
shift introduced by gate noise,
while the infidelity measures the
loss of ground-state population.

Fig.~\ref{fig:robustness}(a) shows the absolute
energy deviations under two-qubit depolarizing
gate noise for \ce{C6H6} [$\mathrm{CAS}(4,4)$],
where $p_{2q}$ denotes the two-qubit
depolarizing noise parameter.
The CD-inspired ansätze retain smaller median
absolute energy deviations than HEA and
SAE-UCCSD over the investigated noise range.
Adaptive-CD-3 and Adaptive-CD-4 exhibit the
lowest median values, with CD-3 performing
comparably. Part of this ordering is inherited
from the noiseless simulations: HEA already has
a relatively large variational error before
noise is introduced.
A different perspective is provided by the
noise-induced deviation
$\delta E_{\mathrm{noise}}$
[Fig.~\ref{fig:robustness}(b)].
Despite its relatively large absolute energy
error, HEA exhibits only a small additional
energy shift under noise. This highlights the
importance of distinguishing total energy
accuracy from noise sensitivity.
A small $\delta E_{\mathrm{noise}}$ does not
necessarily imply an accurate approximation
to the ground-state energy.
By contrast, the CD-inspired circuits combine
more accurate noiseless variational states
with relatively limited noise-induced deviations.
The ground-state infidelity results in
Fig.~\ref{fig:robustness}(c) show a similar
ordering. The adaptive CD circuits exhibit
the lowest median infidelity, followed by CD-3,
indicating that their advantage is not limited
to the energy expectation value. Their noisy
output states also retain a larger population
in the target ground state than those produced
by SAE-UCCSD and HEA.

\begin{figure}[t]
\centering
\includegraphics[width=\linewidth]{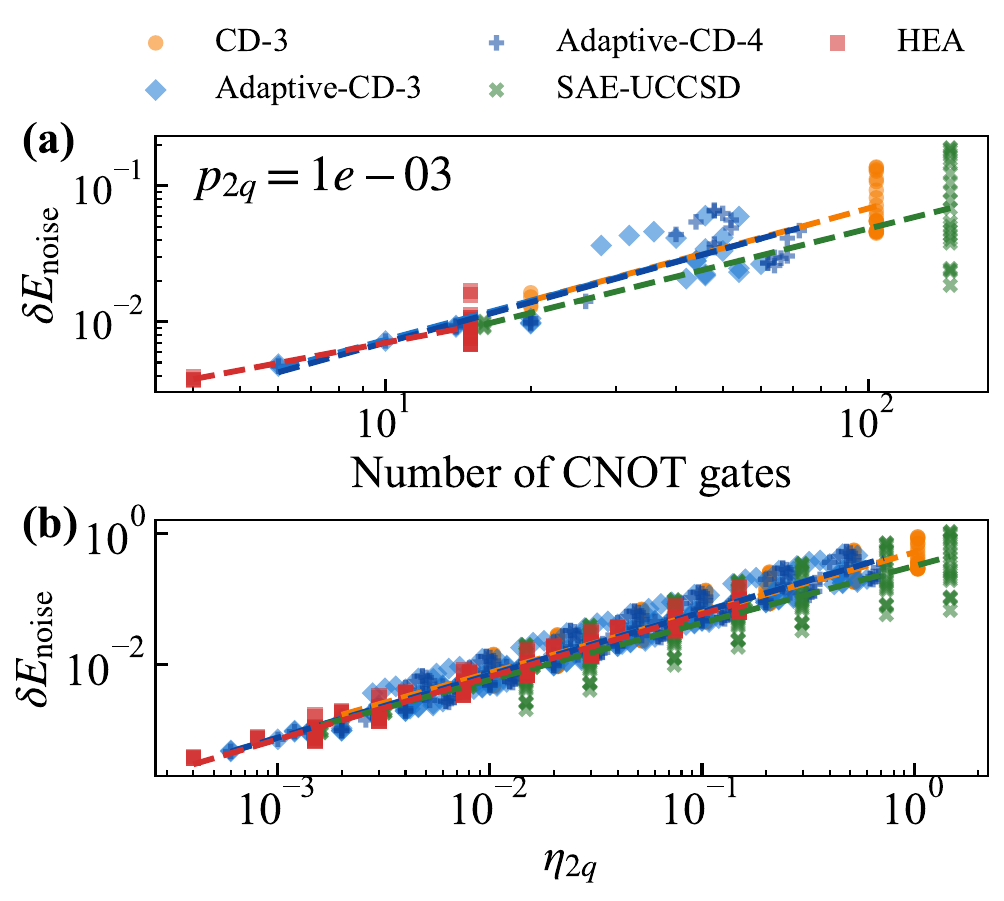}
\vspace{-3mm}
\caption{
Noise-induced energy deviation
$\delta E_{\mathrm{noise}}$~[Eq.~\eqref{eq:noise_error}]
for \ce{C6H6} [$\mathrm{CAS}(4,4)$].
(a) Dependence on the CNOT gate count
$N_{\mathrm{CX}}$ at a fixed two-qubit
error rate $p_{2q}=10^{-3}$.
(b) Dependence on the effective two-qubit
noise exposure
$\eta_{2q}=N_{\mathrm{CX}}p_{2q}$.
Each data point corresponds to a
molecular geometry, variational ansatz,
and noise setting. The dashed line represents the global power-law fit.
}
\label{fig: noise_scaling}
\end{figure}

To relate these observations to circuit
resources, Fig.~\ref{fig: noise_scaling}(a)
plots the noise-induced energy deviation
against the number of CNOT gates at
$p_{2q}=10^{-3}$.
A positive association is observed:
circuits with larger entangling-gate counts
generally exhibit larger noise-induced energy
deviations. This behavior is consistent with
the accumulation of two-qubit gate error, which is usually one of the main sources of error in digital quantum hardware.
At the same time, circuits with similar
CNOT counts can still display different
noise responses, indicating that the detailed
gate sequence and the prepared state also
influence the final energy deviation.

To further quantify the combined effect of
entangling-gate count and depolarizing noise,
we introduce the effective two-qubit noise exposure
$\eta_{2q}\equiv N_{\mathrm{CX}}\times p_{2q}$.
As shown in Fig.~\ref{fig: noise_scaling} (b),
the noise-induced energy deviation for
\ce{C6H6} [$\mathrm{CAS}(4,4)$]
follows an empirical power-law scaling obtained from a global linear regression in logarithmic space,
\begin{equation}
\label{eq:noise_scaling}
\delta E_{\mathrm{noise}}
\propto \eta_{2q}^{0.92},
\qquad R^2\simeq 0.93.
\end{equation}
The exponent is close to unity, suggesting
an approximately linear dependence on the
effective noise exposure over the investigated
range. Although this relation is empirical
and does not capture all circuit-dependent
effects, the relatively large coefficient
of determination indicates that
$\eta_{2q}$ accounts for much
of the observed variation in the
noise-induced energy deviation. These observations motivate the subsequent
hardware benchmarks, where the resource savings
of adaptive CD circuits are assessed after
transpilation to a superconducting processor.

\section{IBM quantum hardware implementation}

\label{sec: exp}

Finally, we evaluate the CD-inspired variational
circuits on superconducting quantum hardware.
Experiments were performed using Qiskit Runtime
on the 156-qubit \texttt{ibm\_basquecountry}
processor, with a nominal shot budget of 8192.
Based on the numerical benchmarks, we focus on
CD-3 and Adaptive-CD-3, which offer favorable
accuracy--resource trade-offs among the
ansätze considered.

\begin{figure}[t]
\centering
\includegraphics[width=\linewidth]{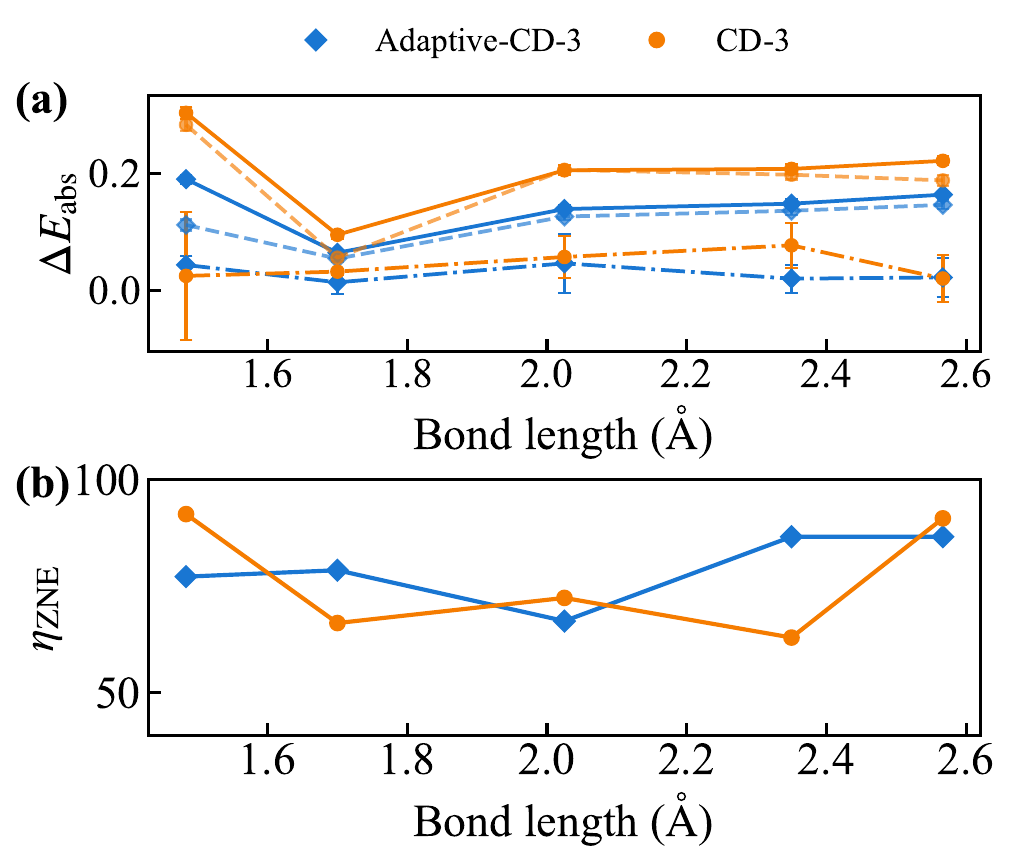}\vspace{-4mm}
\caption{
Hardware benchmarks of CD-3 and
Adaptive-CD-3 for \ce{C6H6}
[$\mathrm{CAS}(4,4)$] on the
\texttt{ibm\_basquecountry} processor.
(a) Absolute energy deviations
$\Delta E_{\mathrm{abs}}$~[Eq.~\eqref{eq:ideal_delta_E}]
across five bond distances.
Error bars indicate the estimator-reported
standard deviations of the energy estimates. Solid, dashed, and dash-dotted lines
correspond to Raw, Readout-mitigated,
and ZNE results, respectively.
(b) Percentage reduction in absolute energy
error achieved by ZNE relative to the Raw configuration, $\eta_{\mathrm{ZNE}}$
[Eq.~\eqref{eq:zne_improvement}].
}
\label{fig: exp}
\end{figure}

We examine three error-mitigation
configurations: (i) without
estimator-level mitigation (Raw),
(ii) with measurement-error mitigation
(Readout), and (iii) with both
measurement-error mitigation and
zero-noise extrapolation
(ZNE)~\cite{Temme2017,Li2017}.
In all configurations, XpXm dynamical
decoupling and gate and measurement twirling
are enabled~\cite{qiskit}. ZNE is implemented through
Qiskit Runtime using gate folding at noise
factors of 1, 3, and 5, with the default
exponential extrapolation procedures.
A nominal budget of 8192 shots per measurement
configuration is used.

The experiments are conducted for \ce{C6H6}
within the four-qubit $\mathrm{CAS}(4,4)$
model at five representative bond distances.
Fig.~\ref{fig: exp}(a) compares the
absolute energy deviations obtained
from hardware measurements, evaluated
relative to the corresponding active-space
FCI reference energies.
Before ZNE,
Adaptive-CD-3 generally exhibits smaller
energy deviations than CD-3. Following ZNE,
the differences between the two ansätze
are reduced at several geometries, while
Adaptive-CD-3 remains competitive
throughout the investigated range.
The resource advantage of adaptive operator
selection is more systematic.
Table~\ref{tab:ibm-resource} summarizes the
circuit resources after instruction set architecture (ISA) transpilation
to the target backend. \SVR{This transpilation step is necessary because IBM quantum processors support only a restricted native gate set and a fixed qubit-coupling topology, so the circuit must be rewritten to satisfy these hardware constraints before execution. For the particular case of \texttt{ibm\_basquecountry}, the native unitary gate set is composed of the one-qubit gates $\mathrm{RZ}(\theta)=\exp(-i\theta Z/2)$, $X$ and $\mathrm{SX}=\sqrt{X}$, together with the controlled-$Z$ (CZ) gate $\mathrm{CZ}=\ket{0}\!\bra{0}\otimes I + \ket{1}\!\bra{1}\otimes Z$.}
Adaptive-CD-3 requires fewer CZ gates than
CD-3 at all five geometries, with reductions
ranging from approximately $30\%$ to $70\%$.
This demonstrates that selecting a subset
of AGP-derived generators can substantially
reduce entangling-gate requirements
even after hardware-aware compilation.

To quantify the improvement in absolute
energy accuracy obtained through ZNE,
we define
\begin{equation}
\label{eq:zne_improvement}
\eta_{\mathrm{ZNE}}
=
100\left(
1-
\frac{
|E_{\mathrm{ZNE}}-E_{\mathrm{FCI}}|
}{
|E_{\mathrm{raw}}-E_{\mathrm{FCI}}|
}
\right),
\end{equation}
where positive values indicate a reduction
in absolute energy error relative to the
Raw configuration.
Fig.~\ref{fig: exp}(b) shows the
relative improvement in energy accuracy
obtained through ZNE for both ansätze.
The effectiveness of ZNE depends on
the molecular geometry and circuit,
emphasizing that error mitigation and
circuit-resource reduction address
different aspects of hardware performance.

\begin{table}[t]
\centering
\caption{
Circuit resources after ISA transpilation
for Adaptive-CD-3 and CD-3 applied to
\ce{C6H6} [$\mathrm{CAS}(4,4)$]
at five molecular geometries.
}
\label{tab:ibm-resource}
\begin{ruledtabular}
\begin{tabular}{lccccc}
$R$ ($\mathrm{\mathring{A}}$) & Ansatz & Depth & RZ & SX & CZ \\
\midrule
\multirow{2}{*}{1.483} & Adaptive-CD-3 & 159 & 72  & 78  & 44  \\
                       & CD-3          & 361 & 176 & 222 & 124 \\
\midrule
\multirow{2}{*}{1.700} & Adaptive-CD-3 & 53  & 28  & 34  & 13  \\
                       & CD-3          & 75  & 107 & 40  & 44  \\
\midrule
\multirow{2}{*}{2.025} & Adaptive-CD-3 & 163 & 80  & 98  & 54  \\
                       & CD-3          & 364 & 171 & 218 & 124 \\
\midrule
\multirow{2}{*}{2.350} & Adaptive-CD-3 & 211 & 111 & 136 & 77  \\
                       & CD-3          & 358 & 168 & 220 & 124 \\
\midrule
\multirow{2}{*}{2.567} & Adaptive-CD-3 & 231 & 115 & 158 & 85  \\
                       & CD-3          & 362 & 169 & 220 & 124 \\
\end{tabular}
\end{ruledtabular}
\end{table}

\section{Discussion and Conclusion}
\label{sec: conclu}

We have investigated resource-efficient 
CD-inspired variational
quantum algorithms for molecular ground-state calculations by combining CAS selection with point-group
symmetry-adapted qubit encoding. While CAS selection restricts the orbital space, symmetry-based encoding further reduces the qubit representation.
Building on previous AGP-derived variational
constructions~\cite{FerreiroVelez2026}
and adaptive CD operator
selection~\cite{Tancara2026}, we construct
CD-inspired operator pools directly from
symmetry-reduced active-space Hamiltonians.
This approach enables the investigation
of fixed and adaptive CD-inspired ans\"atze
for molecular models of \ce{N2},
\ce{C2H4}, and \ce{C6H6}
within compact qubit representations.

Numerical benchmarks demonstrate that
the CD-inspired ansätze achieve competitive
variational accuracy relative to the
corresponding active-space FCI references.
Adaptive operator selection improves
the accuracy of fixed low-weight CD
circuits at selected molecular geometries,
although appreciable deviations remain
in some bond-stretched configurations.
These results highlight the importance of  balancing active-space size, Pauli-weight
truncation, and variational flexibility
when treating more
strongly correlated molecular systems.

Noise simulations reveal the interplay
between intrinsic variational error
and entangling-gate overhead.
Experiments on the
\texttt{ibm\_basquecountry}
superconducting processor further show
that Adaptive-CD-3 requires fewer
two-qubit gates after hardware-aware
transpilation than CD-3
across five geometries of the four-qubit
benzene model, while maintaining
comparable energy accuracy.
{The number of CZ gates is reduced by  approximately 30\% to 70\%, demonstrating that adaptive selection of AGP-derived operators provides circuit-resource savings after hardware-aware compilation.}

Extending this framework to larger
active spaces will require establishing
how AGP-derived operator pools scale
with molecular size and symmetry,
while accounting for the costs of
adaptive selection and energy measurements.
Systematic active-orbital selection,
effective Hamiltonian downfolding,
and hardware-aware operator construction
offer possible directions for further
improvement. The Pauli-rotation structure
also permits implementation across
different gate-based architectures,
including trapped-ion, semiconductor
spin-qubit, and photonic processors
~\cite{Hempel2018,Zhang2026SiliconVQE,
Baldazzi2025}.
These directions provide opportunities
to assess the resource efficiency
of symmetry-reduced CD-inspired VQE
beyond the molecular models and hardware
investigated here.

\section*{Acknowledgments}
This work is partially supported by PID2021-126273NB-I00 and PID2024-157842OA-I00, Projects in the field of Artificial Intelligence 2025 (AIA2025-163435-C44) and the Severo Ochoa Centres of Excellence program through Grant CEX2024-001445-S.
R.Q.X. and J.L.T. appreciate the support from China Scholarship Council (CSC) under Grant Nos. 202206890001 and 202306890004. 
We acknowledge the
use of IBM Quantum services for this work. The views expressed are those of the authors and do not reflect the official
policy or position of IBM or the IBM Quantum team.

\section*{Supporting Information}

Additional active-space benchmarks for
\ce{C2H4} and \ce{N2}, Pauli-weight
truncation analysis for \ce{C6H6},
and detailed definitions of the HEA
and SAE-UCCSD reference ansätze.

\section*{DATA AVAILABILITY}
The code and data, as well as all the results discussed in this work, are available at \url{https://github.com/Ruoqian-Xu/Resource-Efficient-Counterdiabatic-VQE-in-Symmetry-Reduced-Active-Spaces}. 

\bibliography{sample}

\clearpage

\end{document}


\crefname{equation}{Eq.}{Eqs.}
\crefname{figure}{Fig.}{Fig.}
\crefname{appendix}{Appendix}{Appendix}
\renewcommand{\thefigure}{S\arabic{figure}}
\renewcommand{\theequation}{S\arabic{equation}}
\renewcommand{\bibnumfmt}[1]{[S#1]}
\renewcommand{\citenumfont}[1]{S#1}
\renewcommand{\thesection}{S\arabic{section}}

\title{
Supporting Information for: \\ Resource-Efficient Counterdiabatic VQE in Symmetry-Reduced Active Spaces
}

\author{Ruoqian Xu$^{\orcidlink{0009-0005-3242-9674}}$}
\affiliation{Quantum Advanced Research Center (QuARC), CSIC, 28049, Madrid, Spain}
\affiliation{Instituto de Ciencia de Materiales de Madrid (ICMM), CSIC, 28049, Madrid, Spain}
\affiliation{\mbox{Departamento de Física Teórica de la Materia Condensada, Universidad Autónoma de Madrid, 28049 Madrid, Spain}}

\author{Jialiang Tang$^{\orcidlink{0009-0001-6420-6492}}$}
\affiliation{Quantum Advanced Research Center (QuARC), CSIC, 28049, Madrid, Spain}
\affiliation{Instituto de Ciencia de Materiales de Madrid (ICMM), CSIC, 28049, Madrid, Spain}
\affiliation{\mbox{Departamento de Física Teórica de la Materia Condensada, Universidad Autónoma de Madrid, 28049 Madrid, Spain}}

\author{Sebastián V. Romero$^{\orcidlink{0000-0002-4675-4452}}$}
\affiliation{Quantum Advanced Research Center (QuARC), CSIC, 28049, Madrid, Spain}
\affiliation{Instituto de Ciencia de Materiales de Madrid (ICMM), CSIC, 28049, Madrid, Spain}
\affiliation{\mbox{Departamento de Física Teórica de la Materia Condensada, Universidad Autónoma de Madrid, 28049 Madrid, Spain}}

\author{Yue Ban$^{\orcidlink{0000-0003-1764-4470}}$}
\email{yue.ban@csic.es}
\affiliation{Quantum Advanced Research Center (QuARC), CSIC, 28049, Madrid, Spain}
\affiliation{Instituto de Ciencia de Materiales de Madrid (ICMM), CSIC, 28049, Madrid, Spain}

\author{Xi Chen$^{\orcidlink{0000-0003-4221-4288}}$}
\email{xi.chen@csic.es}
\affiliation{Quantum Advanced Research Center (QuARC), CSIC, 28049, Madrid, Spain}
\affiliation{Instituto de Ciencia de Materiales de Madrid (ICMM), CSIC, 28049, Madrid, Spain}
    
\maketitle
\tableofcontents

\begin{figure}[t]
\centering
\includegraphics[width=\linewidth]%
{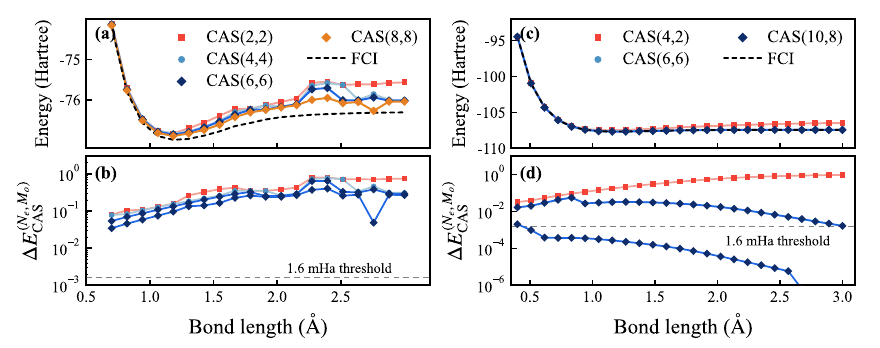}
\caption{
Active-space dependence of the potential-energy
curves and corresponding energy deviations
for \ce{C2H4} and \ce{N2}.
(a) and (c) Potential-energy curves for
\ce{C2H4} and \ce{N2}, respectively.
(b) and (d) Corresponding absolute energy
deviations relative to the full-space
FCI reference~[Eq.~\eqref{eq:cas_deviation_appendix}].
The active spaces considered are
CAS(2,2), CAS(4,4), CAS(6,6), and
CAS(8,8) for \ce{C2H4}, and
CAS(4,2), CAS(6,6), and CAS(10,8)
for \ce{N2}.
The horizontal dashed lines in
(b) (d) indicate the $1.6$~mHa
energy-deviation threshold.
}
\label{fig:cas}
\end{figure}

\section{Active-Space Dependence}
\label{app:cas_comparison}

To complement the benzene active-space analysis in Section II.B of the main text,, we examine
the dependence of the potential-energy curves
of \ce{C2H4} and \ce{N2} on the selected
active spaces. The energy deviation relative
to the full-space FCI reference is defined as
\begin{equation}
\label{eq:cas_deviation_appendix}
\Delta E_{\mathrm{CAS}}^{(N_{\mathrm e},M_{\mathrm o})}(R)
=
\left|
E_{\mathrm{CASCI}(N_{\mathrm e},M_{\mathrm o})}(R)
-
\widetilde{E}_{\mathrm{FCI}}(R)
\right|,
\end{equation}
where $N_{\mathrm e}$ and $M_{\mathrm o}$
denote the numbers of active electrons
and spatial orbitals, respectively.
Here, $\widetilde{E}_{\mathrm{FCI}}(R)$
denotes the full-space FCI energy in
the same finite one-electron basis
used for the CASCI calculations.

For \ce{C2H4}, we compare
$\mathrm{CAS}(2,2)$, $\mathrm{CAS}(4,4)$,
$\mathrm{CAS}(6,6)$, and $\mathrm{CAS}(8,8)$~\cite{D3SC02516K}.
As shown in Fig.~\ref{fig:cas}(a),
all four models reproduce the qualitative
shape of the potential-energy curve.
Increasing the active space beyond
$\mathrm{CAS}(4,4)$ does not produce a
uniform improvement relative to the
plotted reference over the investigated
geometries. We therefore adopt
$\mathrm{CAS}(4,4)$ for the VQE benchmarks,
while recognizing that quantitative
electronic-structure accuracy may require
a more suitable selection of active orbitals.

For \ce{N2}, Fig.~\ref{fig:cas}(b)
compares $\mathrm{CAS}(4,2)$,
$\mathrm{CAS}(6,6)$, and
$\mathrm{CAS}(10,8)$~\cite{dcunha2024state,sugisaki2022adiabatic}.
The $\mathrm{CAS}(6,6)$ model
provides a substantially improved
potential-energy profile relative to
$\mathrm{CAS}(4,2)$, whereas the
additional changes obtained with
$\mathrm{CAS}(10,8)$ are comparatively
modest in the investigated range.
We consequently use $\mathrm{CAS}(6,6)$
in the main calculations.

\begin{figure}[t]
\centering
\includegraphics[width=0.7\linewidth]{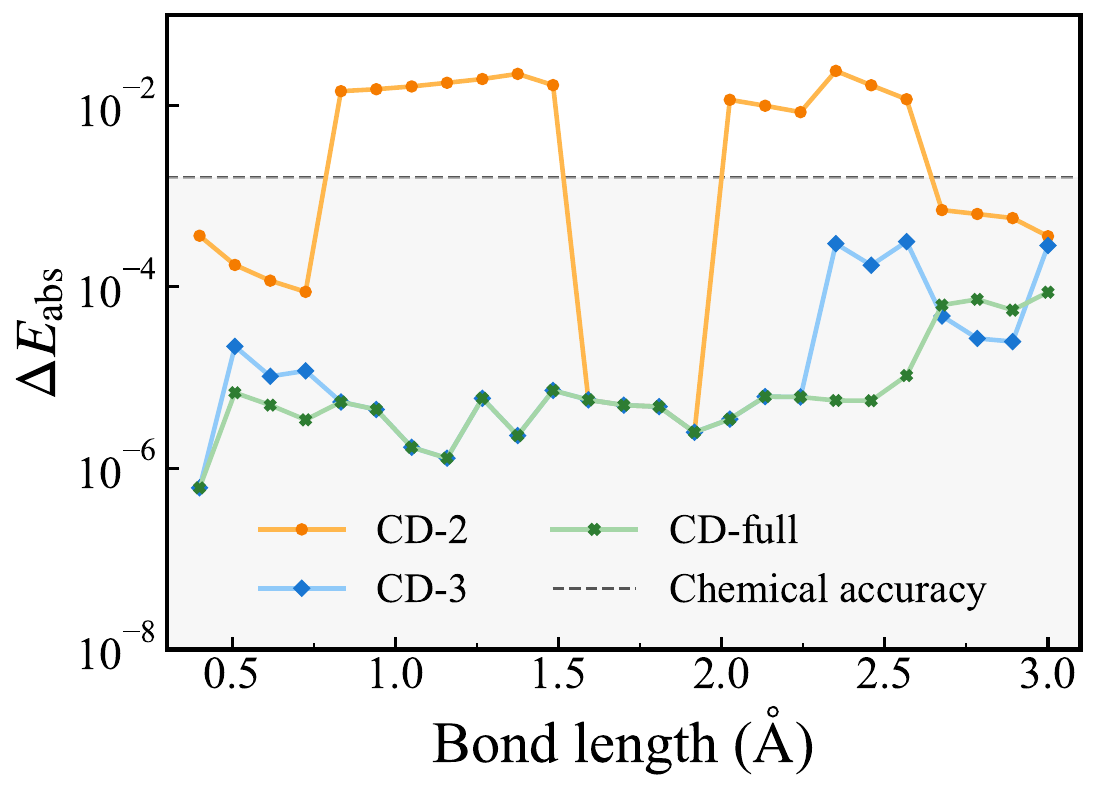}\vspace{-3mm}

\caption{
Effect of Pauli-weight truncation
on the variational energy accuracy
for \ce{C6H6} [CAS(4,4)].
Absolute energy deviations
$\Delta E_{\mathrm{abs}}
=|E-E_{\mathrm{FCI}}|$
for CD-2, CD-3, and CD-full
as functions of bond length,
where $E_{\mathrm{FCI}}$ is the exact
ground-state energy of the corresponding
active-space Hamiltonian.
The horizontal dashed line marks
the $1.6$~mHa accuracy threshold.
}
\label{fig: cd_truncation}
\end{figure}

\section{Reference Variational Ansätze}
\label{ansatz}

As a hardware-oriented benchmark, we employ
a HEA consisting
of parameterized single-qubit rotations
and nearest-neighbor CNOT gates.
Starting from the encoded HF state,
the circuit is constructed from repeated
layers of
\begin{align}
U_{\mathrm{rot}}^{(l)}
&=
\bigotimes_{q=0}^{n_q-1}
R_z(\phi_{l,q})R_y(\theta_{l,q}),
\\
U_{\mathrm{ent}}^{(l)}
&=
\prod_{q=0}^{n_q-2}
\mathrm{CNOT}_{q\rightarrow q+1}.
\end{align}
We use $D=2$ layers, with
$2n_qD$ variational parameters.
The gate ordering is kept fixed
throughout the calculations. In addition, the UCCSD~\cite{romero2018strategies} is used as a chemistry-motivated benchmark.
Its wavefunction is
\begin{equation}
|\Psi_{\mathrm{UCCSD}}\rangle
=
e^{\hat T-\hat T^\dagger}
|\Phi_{\mathrm{HF}}\rangle,
\end{equation}
where $\hat T=\hat T_1+\hat T_2$,
with
\begin{align}
\hat T_1
&=
\sum_{i\in\mathrm{occ}}
\sum_{a\in\mathrm{virt}}
t_i^a a_a^\dagger a_i,
\\
\hat T_2
&=
\frac{1}{4}
\sum_{\substack{
i,j\in\mathrm{occ}\\
a,b\in\mathrm{virt}}}
t_{ij}^{ab}
a_a^\dagger a_b^\dagger a_j a_i.
\end{align}
Here, $i,j$ and $a,b$ denote occupied
and virtual spin orbitals, respectively,
and the cluster amplitudes are
variational parameters.

In the present calculations, the fermionic
excitation operators are generated using
Qiskit Nature and subsequently mapped
into the symmetry-reduced active-space
representation using the QuantumSymmetry
encoding. The resulting ansatz is denoted
SAE-UCCSD.

 \section{Pauli-Weight Truncation}
\label{app: cd truncation}

To examine the effect of Pauli-weight
truncation, we compare CD-2, CD-3,
and CD-full for the four-qubit
$\mathrm{CAS}(4,4)$ model of \ce{C6H6}.
CD-2 and CD-3 retain AGP-derived
Pauli strings with maximum weights
of two and three, respectively,
whereas CD-full includes the
complete operator pool generated
by the chosen AGP approximation.

Fig.~\ref{fig: cd_truncation}
shows the absolute energy deviations $\Delta E_{\text{abs}}$
relative to the active-space FCI
reference over the investigated
bond-distance range.
CD-2 exhibits larger errors at
several geometries, whereas CD-3
achieves accuracy comparable to
CD-full over much of the range.
The additional weight-four
operators therefore provide only
limited improvements for the
considered four-qubit Hamiltonian,
supporting the use of CD-3
in the main benchmarks.


\bibliographystyle{unsrt}
\bibliography{sample}